\documentclass{aastex}
\usepackage{spr-astr-addons}
\usepackage{url}

\RequirePackage{color}

\newcommand{\emaila}{rsharma@associates.iucaa.in}

\begin{document}

\title{Anisotropic extension of Finch and Skea stellar model}
\shorttitle{Anisotropic extension of Finch and Skea}
\shortauthors{Sharma {\em et al}}

\author{Ranjan Sharma} \and \author{Shyam Das}
\affil{Department of Physics, P. D. Women's College, Jalpaiguri, India.}
\and
\author{S. Thirukkanesh}
\affil{Department of Mathematics, Eastern University, Chenkalady, Sri Lanka.}
\email{\emaila}

\begin{abstract}
In this paper, the spacetime geometry of Finch and Skea [{\it Class. Quantum Grav.} {\bf6} (1989) 467] has been utilized to obtain closed-form solutions for a spherically symmetric anisotropic matter distribution. By examining its physical admissibility, we have shown that the class of solutions can be used as viable models for observed pulsars. In particular, a specific class of solutions can be used as an `anisotropic switch' to examine the impact of anisotropy on the gross physical properties of a stellar configuration. Accordingly, the mass-radius relationship has been analyzed. 
\end{abstract}

\keywords{Finch and Skea ansatz; Exact solutions; Anisotropic star}

\section{Introduction}
The spacetime geometry of \cite{Finch89}, originally developed by \cite{Duorah87}, has got much attention in the modelling of relativistic compact stars as the solution is well-behaved and satisfies all criteria of physical acceptability (\cite{Delgaty98}). The Finch-Skea isotropic stellar model has subsequently been generalized by many investigators to study a large variety of stellar bodies by incorporating electro-magnetic field and anisotropic pressure. The Finch and Skea ansatz has also been utilized to interpret astrophysical systems in lower as well as in higher dimensional gravitational theories. 

Recently, by making use of the Finch-Skea ansatz, \cite{Maharaj16} have developed new families of exact solutions for an anisotropic charged matter distribution and showed that the masses and radii obtained for the class of solutions are consistent with observational data (\cite{Kileba17}). In this approach, by specifying the metric function $g_{rr}$, the system is solved for the other metric function $g_{tt}$ in terms of the Bessel functions which eventually gets transformed to simple trigonometric/algebraic forms for a specific range of parameter values. The original Finch-Skea stellar model as well as the charged stellar model developed by \cite{Hansraj06} can be regained from the new class of solutions. For a specific charge distribution, \cite{Ratanpal17} obtained a new class of solutions in terms of Bessel functions for a matter distribution in the Finch-Skea background spacetime. The physically viable solution has been utilized to study the impact of charge on the mass-radius ($M-R$) relationship, in particular. Exact solutions in terms of Bessel functions for spherically symmetric anisotropic systems in Finch-Skea background spacetime was also obtained by \cite{Sharma2013}. Earlier, \cite{Pandya15} presented a new class of solutions for a spherically symmetric anisotropic matter distribution by making a generalization of the Finck-Skea ansatz and showed that their class of solutions is compatible with a wide variety of observed compact stars.  The Finch-Skea ansatz has been used by \cite{Sharma13} to develop an anisotropic stellar model which has been shown to admit a quadratic EOS. The solution is, in fact, a sub-class of the model developed by \cite{Pandya15}.  \cite{Tikekar07} have shown that the Finch-Skea model can be used to describe ultra-compact stars like `strange stars' composed of $u$, $d$ and $s$ quarks. 

In ($2+1$) dimensions, the Finch-Skea ansatz has been utilized by \cite{Bhar14} to obtain interior solutions corresponding to the BTZ (\cite{BTZ}) exterior spacetime for a matter distribution obeying the MIT Bag model (\cite{MIT}) equation of state (EOS). The Finch-Skea stellar model in $(2+1)$ dimensions has been analyzed by \cite{Ayan13}. The Finch-Skea ansatz has also been taken up in higher dimensional gravitational theories (\cite{Hansraj16,Dadhich16,Molina17}). In a recent paper, assuming the Finch-Skea ansatz as a seed solution, \cite{Hansraj17} have constructed a model of a static spherical distribution of perfect fluid in trace-free Einstein gravity.

The objective of the current investigation is to generate exact interior solutions corresponding to the Schwarzschild exterior spacetime which can be utilized to describe realistic stars. Since we are interested in studying highly compact stars, we intend to obtain solutions for an anisotropic matter distribution whose geometry will be characterized by the Finch and Skea ansatz. The reason for incorporating anisotropy is due to the fact in the high-density regime of compact stars the radial pressure $p_r$ and the transverse pressure $p_t$ need not be equal (\cite{Ruderman72,Canuto74}). Origin of pressure anisotropy in stellar objects has been extensively analyzed by \cite{Bowers74}. Some of the reasons attributed to the origin of anisotropic stress within a stellar distribution are due to the existence of a solid core (\cite{Kippen}), the presence of electromagnetic field (\cite{Weber,Usov}), phase transition (\cite{Sokolov}), pion condensation (\cite{Sawyer}) etc. Scalar field `boson stars' are naturally anisotropic (\cite{Schunck}). Wormholes (\cite{Morris}) and gravastars (\cite{Cattoen,DeBend}) have also anisotropic characteristics. \cite{Ivanov10} has pointed out that incorporation of the electromagnetic field into a relativistic stellar  object has an anisotropic interpretation. In a recent paper, \cite{Ivanov17} has analyzed the different class of anisotropic models developed so far resembling the charged isotropic solutions. 

In this work, the Finch and Skea stellar model has been extended to the case of an anisotropic matter distribution. The system of field equations has been solved rigorously to generate analytic solutions which are physically viable. The paper has been organized as follows: In Section \ref{sec1}, the Einstein field equations for a static spherically symmetric and anisotropic fluid distribution have been laid down. By making use to the \cite{Durga} transformation equations, an equivalent set of field equations have been obtained. In Section \ref{sec2}, for a particular anisotropic profile, we have provided three different class of solutions. The solutions have been matched to the Schwarzschild exterior metric at the boundary for the three cases in Section \ref{sec3}. In Section \ref{sec4}, by satisfying relevant physical requirements, we have shown that the model can accommodate some observed pulsars. We have also critically analyzed the impact of anisotropy on the gross physical behaviour of a compact star for our class of solutions. We have concluded by discussing our results in Section \ref{sec5}.

\section{The field equations}\label{sec1}
To describe the gravitational field of the interior of a static and spherically symmetric relativistic stellar configuration, we write the line element in coordinates $(x^{a}) = (t,r,\theta,\phi)$ as
\begin{equation}
\label{1} ds^{2} = -e^{2\nu(r)} dt^{2} + e^{2\lambda(r)} dr^{2} +
 r^{2}(d\theta^{2} + \sin^{2}{\theta} d\phi^{2}).
\end{equation}
The energy momentum tensor for an anisotropic fluid distribution is assumed to be of the form
\begin{equation}
T_{ij} = (\rho + p_t)u_{i} {u_j} + p_{t} g_{ij} + (p_r - p_t)\chi_{i} \chi_{j}.\label{2}
\end{equation}
The energy density $\rho$, the radial pressure $p_r$ and  the tangential pressure $p_t$ are measured relative to the comoving
fluid velocity $u^i = e^{-\nu}\delta^i_0.$ In Eq.~(\ref{2}), $\chi^i$ is a unit space-like $4$-vector along the radial direction.
 For the line element
(\ref{1}) and matter distribution (\ref{2}), the Einstein field
equations can be expressed as
\begin{eqnarray}
\label{3} \rho &=& \frac{1}{r^{2}} \left[ r(1-e^{-2\lambda})
\right]',\\
\label{4} p_r &=& - \frac{1}{r^{2}} \left( 1-e^{-2\lambda} \right)
+
\frac{2\nu'}{r}e^{-2\lambda} ,\\
\label{5}p_t &=& e^{-2\lambda}\left( \nu'' + \nu'^{2} +
\frac{\nu'}{r}- \nu'\lambda' - \frac{\lambda'}{r} \right) ,
\end{eqnarray}
where a prime $(')$ denotes differentiation with respect to $r$. The
field Eqs.~(\ref{3})-(\ref{5}) have been written in system of units having $8\pi G = 1 = c$. A different but equivalent form of the field equations can be found if we introduce the transformation first proposed by \cite{Durga}
\begin{equation}
\label{7} x = \frac{r^2}{R^2},~~ Z(x)  = e^{-2\lambda(r)} ~\mbox{and}~
A^{2}y^{2}(x) = e^{2\nu(r)},
\end{equation}
where $A$ and $R$ are constants.  Under the transformation (\ref{7}), the system of Eqs.~(\ref{3})-(\ref{5}) take the form
\begin{eqnarray}
\label{9}   \rho &=& \frac{1-Z}{x R^2} - \frac{2\dot{Z}}{R^2}, \\
\label{10} p_r &=& 4Z\frac{\dot{y}}{yR^2} + \frac{Z-1}{xR^2}, \\
\label{11} p_t &=& p_r +\Delta, \\
\label{12}0 & =& 4x^2 Z\ddot{y}+ 2x^2 \dot{Z}\dot{y} + \nonumber\\
&&\left(x
\dot{Z} -Z+1-\Delta x R^2\right)y,
\end{eqnarray}
where $\Delta = p_t-p_r$ is the measure of anisotropy and a dot $(.)$ denotes differentiation with respect to the variable $x$. The line element  (\ref{1})  now gets the form
\begin{equation}
\label{8} ds^2 = -A^2 y^2 dt^2 + \frac{R^2}{4xZ}dx^2 + x R^2(d\theta^2
+\sin^2\theta d\phi^2).
\end{equation}
The total mass contained within a radius $r$ 
\begin{equation}
\label{6} m(r)= \frac{1}{2}\int_0^r\tilde{r}^2 \rho(\tilde{r})d\tilde{r},
\end{equation}
in terms of the new variables takes the form
\begin{equation}
\label{13}m(x)=\frac{1}{4} \int_0^x\sqrt{\omega}\rho(\omega)d\omega.
\end{equation}
 
\section{Generating exact solutions}\label{sec2}
We seek solutions to the system comprising four equations (\ref{9})-(\ref{12}) in six unknowns $Z, y, \rho, p_r, p_t$ and
$\Delta$. Eq.~(\ref{12}) turns out to be the  master equation in  the integration process. To integrate Eq.~(\ref{12}), we specify the
gravitational potential $Z$ and the measure of anisotropy $\Delta$ on regularity and physical grounds.  We make the following choices
\begin{eqnarray}
\label{14} Z &=& \frac{1}{1+x}, \\
\label{15} \Delta &=& \frac{\alpha  x}{R^2(1+ x)^2},
\end{eqnarray}
where the constant $\alpha$ is the measure of anisotropy. The $\alpha =0$ case corresponds to the well-known Finch-Skea model.
Note that the choice in Eq.~(\ref{14}) is non-singular at the origin and was previously used to study neutron stars with isotropic matter distribution by \cite{Finch89}. Our choice of $\Delta$ is reasonable for the following reasons: (i) $\Delta$ is regular at the centre (i.e., $p_r=p_t$ at the centre), (ii) provides a wide range of anisotropy and (iii) the particular choice makes the system of equations integrable.

With these assumptions we obtain 
\begin{equation}
\rho=\frac{3+\frac{r^2}{R^2}}{R^2(1+\frac{r^2}{R^2})^2},\label{den}
\end{equation} 
\begin{equation}
m(r)=\frac{r^3}{2(r^2+R^2)},\label{mass}
\end{equation} 
\begin{equation}
\Delta=\frac{\alpha r^2}{R^4(1+\frac{r^2}{R^2})^2}. \label{ani1}
\end{equation} 
To evaluate the remaining physically interesting quantities, we need to determine $y(r)$ which can be obtained in the following way. 

Substitution of Eqs.~(\ref{14})  and (\ref{15}) in Eq.~(\ref{12}) yields
\begin{equation}
\label{16} 4(1+x) \ddot{y}-2 \dot{y}+ (1-\alpha)y = 0.
\end{equation}
It is noteworthy that equation (\ref{16}) can be obtained as a special case by setting $a=1$ in equation ($10$) of the paper by \cite{ Maharaj16}. This is so because, in our work, we have assumed $x=r^2/R^2$ and $Z=1/(1+x)$ whereas in the formulation of \cite {Maharaj16} the following assumptions were made: $x=C r^2$, $Z=1/(1+ax)$. We must point out here that in the studies of \cite{Maharaj16}, a large family of exact solutions to the master equation in terms of elementary functions, Bessel functions and modified Bessel functions have been generated.  It has also been illustrated with several examples that it is possible to express the solutions in terms of elementary functions and special functions by suitably fixing the model parameters. Subsequently, the solutions have been utilized to model realistic stars (\cite{ Kileba17}). 

In this work, by introducing the transformation
\begin{equation}
\label{17} 1+x  = X,~~ y(x)= Y(X), 
\end{equation}
we rewrite the master equation Eq.~(\ref{16}) in the form
\begin{equation}
\label{18} 4 X \frac{d^{2}Y}{dX^{2}}- 2\frac{dY}{dX} + (1 -\alpha
)Y = 0.
\end{equation}
It should be emphasized here that the transformation (\ref{17}) is crucial in our approach as it leads to Eq.(\ref{18}) in which $X=0$ is a regular singular point. Accordingly, the method of Frobenius can be applied to generate a series solution of the equation which eventually can be expressed in terms of elementary functions. To illustrate this, we first write the solution of the differential Eq.~(\ref{18}) in the series form 
\begin{equation}
\label{19} Y = \sum_{n=0}^{\infty}c_{n}X^{n+s}, ~~ c_{0}\not=0
\end{equation}
where $c_{n}$ are the coefficients of the series and $s$ is a
constant. Substituting Eq.~(\ref{19}) in the differential Eq.~(\ref{18}), we have
\begin{eqnarray}
\label{20}
\sum_{n=1}^{\infty} \left[2ac_{n+1}(n+s+1)[2(n+s)-1]+c_{n}(1-\alpha)\right]X^{n+s}\nonumber\\
+2c_{0}s[2(s-1)-1]X^{s-1}=0. 
\end{eqnarray}
For consistency the coefficients of the various powers of $X$ must
vanish in Eq.~(\ref{20}). Equating the coefficient of $X^{s-1}$ in Eq.~(\ref{20}) to zero, we get
\[ 2c_{0}s[2(s-1)-1] = 0, \]
which is the indicial equation.  Since $c_{0}\not=0$,  we must
have $s=0$ or $s= \frac{3}{2}.$
 Equating the coefficient of $X^{n+s}$ in Eq.~(\ref{20}) to zero we obtain
\begin{equation}
\label{21} c_{n+1}  =  \frac{(\alpha -1)}{2(n+1+s)
[2(n+s)-1]}c_{n} , ~~ n\geq 0. \\
\end{equation}
The result in (\ref{21}) is the basic equation which
determines the nature of the solution.

We can establish a general structure for all the coefficients by
considering the leading terms. We note that the coefficients $c_1,
c_2, c_3, ...$ can all be written in terms of the leading
coefficient $c_0$ and this leads to the expression
\begin{equation}
\label{22} c_{n+1} = \prod_{p=0}^{n}\frac{(\alpha
-1)}{2(p+1+s)[2(p+s)-1]}c_{0}.
\end{equation}
It is also
possible to establish  the result (\ref{22}) rigorously by using
the principle of mathematical induction. We can now generate two
linearly independent solutions from Eqs.~(\ref{19}) and (\ref{22}). For the parameter value $s=0$, we obtain the first solution
\begin{eqnarray}
Y_{1}& =& c_{0}
\left[1+\sum_{n=0}^{\infty}\prod_{p=0}^{n}\frac{(\alpha -1)}{2(p+1)(2p-1)}X^{n+1} \right], \nonumber\\
\label{23} y_{1} &=& c_{0}
\left[1+\right.\nonumber\\
&&\left.\sum_{n=0}^{\infty}\prod_{p=0}^{n}\frac{(\alpha
-1)}{2(p+1)(2p-1)} (1+x)^{n+1} \right].
\end{eqnarray}
For the parameter value $s=\frac{3}{2}$, we obtain the second
solution
\begin{eqnarray}
 Y_{2} &=& c_{0}X^{\frac{3}{2}}
\left[1+\sum_{n=0}^{\infty} \prod_{p=0}^{n}\frac{(\alpha
-1)}{(2p+5)(2p+2)}X^{n+1}
\right], \nonumber \\
\label{24} y_{2}& =& c_{0}(1+x)^{\frac{3}{2}}
\left[1+\right.\nonumber\\
&&\left.\sum_{n=0}^{\infty}\prod_{p=0}^{n}\frac{(\alpha
-1)}{(2p+5)(2p+2)}(1+x)^{n+1} \right].
\end{eqnarray}
Therefore, the general solution to the differential Eq.~(\ref{16}) is given by
\begin{equation}
\label{25} y = a_1 y_{1}(x) + b_1y_{2}(x),
\end{equation}
where $a_1$ and $b_1$ are arbitrary constants and $y_{1}$ and
$y_{2}$ are given by Eqs.~(\ref{23}) and (\ref{24}), respectively. It is interesting to note that the series solution can be written
in terms special functions as demonstrated below:

\subsection{Case I: $0 \leq \alpha < 1$}
From Eq.~(\ref{23}), we have
\begin{eqnarray}
y_1 =  c_0 \left[ 1 + \sum_{n=0}^{\infty}\prod_{p=0}^{n}
\frac{-(1-\alpha)}{2 (p+1)(2p-1)}(\sqrt{1+x})^{2n+2}\right] \nonumber\\
=c_0 \left(\left[1 - \frac{(\sqrt{(1-\alpha)(1+x)})^2}{2!}
+ \frac{(\sqrt{(1-\alpha)(1+x)})^4}{4!} \right.\right.\nonumber \\
\left.- \frac{(\sqrt{(1-\alpha)(1+x)})^6}{6!} +...\right]\nonumber\\
+ \sqrt{(1-\alpha)(1+x)}\left[\sqrt{(1-\alpha)(1+x)}\right.\nonumber\\
\left. \left.-\frac{(\sqrt{(1-\alpha)(1+x)})^3}{3!}+
 \frac{(\sqrt{(1-\alpha)(1+x)})^5}{5!}-...
 \right]\right)\nonumber\\
= c_0 \cos{\sqrt{(1-\alpha)(1+x)}} + \nonumber\\
c_0\sqrt{(1-\alpha)(1+x)}
 \sin{\sqrt{(1-\alpha)(1+x)}}. \nonumber
 \end{eqnarray}
From Eq.~(\ref{24}), we have
\begin{eqnarray}
y_2&= &c_0 (\sqrt{1+x})^3\times\nonumber\\
&&\left[ 1 +
\sum_{n=0}^{\infty}\prod_{p=0}^{n}
\frac{-(1-\alpha)}{(2p+5)(2p+2)}(\sqrt{1+x})^{2n+2}\right] \nonumber\\
&=& \frac{3c_0}{(\sqrt{1-\alpha})^3} \left( \left[
\sqrt{(1-\alpha)(1+x)}-\frac{(\sqrt{(1-\alpha)(1+x)})^3}{3!}
\right. \right. \nonumber\\
& & \left. + \frac{(\sqrt{(1-\alpha)(1+x)})^5}{5!}- ... \right]
 - \sqrt{(1-\alpha)(1+x)} \left[ 1 \frac{}{}  \right. \nonumber\\
 & &  \left.\left. - \frac{(\sqrt{(1-\alpha)(1+x)})^2}{2!}+
 \frac{(\sqrt{(1-\alpha)(1+x)})^4}{4!}
 - ...\right]\right) \nonumber\\
 &=& \frac{3 c_0}{(\sqrt{1-\alpha})^3}
\left[\sin\sqrt{(1-\alpha)(1+x)} -\right.\nonumber\\
&&\left.\sqrt{(1-\alpha)(1+x)}\cos\sqrt{(1-\alpha)(1+x)}\right].\nonumber
\end{eqnarray}
Consequently, the general solution takes the form 
\begin{eqnarray}
 y = \left[D_1 - D_2
\sqrt{(1-\alpha)(1+x)}\right] \cos\sqrt{(1-\alpha)(1+x)} \nonumber\\
\label{26}  + \left[D_2 + D_1
\sqrt{(1-\alpha)(1+x)}\right]\sin\sqrt{(1-\alpha)(1+x)},
\end{eqnarray}
where $D_1$ and $D_2$ are new arbitrary constants. 

The radial and tangential pressure are obtained as 
\begin{equation}
p_r= \frac{F_1 \cos \xi+ F_2\sin \xi}{(r^2+R^2)[(D_1-D_2 \xi)\cos \xi+(D_2+D_1 \xi)\sin \xi]}, \label{pr1}
\end{equation} 
\begin{equation}
p_t= \frac{F_3\cos \xi+F_4\sin \xi}{(r^2+R^2)^2[(D_1-D_2 \xi)\cos \xi+(D_2+D_1 \xi)\sin \xi]}, \label{pt1}
\end{equation} 
where $$\xi= \sqrt{\left(1+\frac{r^2}{R^2}\right)(1-\alpha)},~\xi_1 =\xi^2-\alpha, ~\xi_2=\xi^2+\alpha,$$
$$F_1=(D_1+D_2 \xi-2 \alpha D_1),~ F_2=(D_2-D_1 \xi-2 \alpha D_2),$$
$$F_3=(D_2 \xi R^2\xi_2+D_1\xi_1 R^2),~F_4=(D_2 \xi_1 R^2-D_1 R^2\xi_2 \xi).$$

Note that for $\alpha=0$, we regain the \cite{Finch89} solution
\begin{eqnarray}
\label{26a} y&=&\left[D_1 - D_2 \sqrt{1+x}\right]\cos\sqrt{1+x}\nonumber\\
&&+\left[D_2 + D_1 \sqrt{1+x}\right]\sin\sqrt{1+x},
\end{eqnarray}
for a neutron star with isotropic pressure.

\subsection{Case II: $\alpha = 1$}
In this particular case, from Eq.~(\ref{25}), we have
\begin{equation}
\label{27} y= a_1+b_1 (1+x)^{\frac{3}{2}},
\end{equation}
\begin{equation}
p_r= -\frac{1}{r^2+R^2}+\frac{6 b_1 R^2}{a_1\sqrt{(r^2+R^2)}+b_1(r^2+R^2)^2}, \label{pr2}
\end{equation} 
\begin{equation}
p_t=\frac{r^2(\alpha-1)-R^2}{(r^2+R^2)^2}+ \frac{6 b_1 R^2}{a_1\sqrt{(r^2+R^2)}+b_1(r^2+R^2)^2}. \label{pt2}
\end{equation}

\subsection{Case III: $\alpha > 1$}
In this case, from Eq.~(\ref{23}), we obtain
\begin{eqnarray}
y_1=& c_0 \left[ 1 + \sum_{n=0}^{\infty}\prod_{p=0}^{n}
\frac{(\alpha -1)}{2 (p+1)(2p-1)}(\sqrt{1+x})^{2n+2}\right] \nonumber\\
 =& c_0 \left(\left[1 + \frac{(\sqrt{(\alpha-1)(1+x)})^2}{2!}\right.\right.\nonumber\\
&\left.+ \frac{(\sqrt{(\alpha -1)(1+x)})^4}{4!} +\frac{(\sqrt{(\alpha -1)(1+x)})^6}{6!}+...\right]\nonumber\\
&- \sqrt{(\alpha -1)(1+x)}\left[\sqrt{(\alpha -1)(1+x)} \right.\nonumber\\
& \left. \left. +\frac{(\sqrt{(\alpha -1)(1+x)})^3}{3!}+
 \frac{(\sqrt{(\alpha -1)(1+x)})^5}{5!}+...                                                                         
 \right]\right)\nonumber \\      
=& c_0 \cosh{\sqrt{(\alpha -1)(1+x)}} \nonumber\\
&- c_0 \sqrt{(\alpha -1)(1+x)}
  \sinh{\sqrt{(\alpha
 -1)(1+x)}}. \nonumber
 \end{eqnarray}
From Eq.~(\ref{24}), we have
\begin{eqnarray}
y_2=&c_0 (\sqrt{1+x})^3\left[ 1 +\right.\nonumber\\
&\left.\sum_{n=0}^{\infty}\prod_{p=0}^{n}
\frac{(\alpha -1)}{(2p+5)(2p+2)}(\sqrt{1+x})^{2n+2}\right] \nonumber\\
=& \frac{- 3c_0}{(\sqrt{\alpha -1})^3} \left( \left[
\sqrt{(\alpha -1)(1+x)}+ \frac{(\sqrt{(\alpha -1)(1+x)})^3}{3!}\right. \right.\nonumber\\
 &\left. + \frac{(\sqrt{(\alpha -1)(1+x)})^5}{5!}+ ... \right]-
\sqrt{(\alpha -1)(1+x)} \left[ 1 \right. \nonumber\\
&\left. \left.+ \frac{(\sqrt{(\alpha -1)(1+x)})^2}{2!} +
\frac{(\sqrt{(\alpha -1)(1+x)})^4}{4!} + ...\right]\right) \nonumber\\
=& \frac{- 3c_0}{(\sqrt{\alpha
-1})^3}(\sinh\sqrt{(\alpha-1)(1+x)} -\nonumber\\
&\sqrt{(\alpha-1)(1+x)}\cosh\sqrt{(\alpha-1)(1+x)}), \nonumber
\end{eqnarray}
so that the general takes the final form
\begin{eqnarray}
 y=\left[D_4 - D_3
\sqrt{(\alpha -1)(1+x)}\right] \sinh\sqrt{(\alpha -1)(1+x)} \nonumber\\
+ \left[D_3-D_4 \sqrt{(\alpha
-1)(1+x)}\right]\cosh\sqrt{(\alpha -1)(1+x)},\label{eq:b27}
\end{eqnarray}
where $D_3$ and $D_4$ are new arbitrary constants. In this case, we obtain 
\begin{equation}
p_r= \frac{F_5 \cosh \psi+ F_6\sinh \psi}{(r^2+R^2)[F_9\cosh \psi+F_10\sinh \psi]}, \label{pr3}
\end{equation} 
\begin{equation}
p_t=\frac{F_7\cosh \psi+F_8\sinh \psi}{(r^2+R^2)^2[F_9\cosh \psi+F_10\sinh \psi]} ,\label{pt3}
\end{equation} 
where $$ \psi^2=- \xi^2,~\psi_1=\alpha-\psi^2,~\psi_2=\alpha+\psi^2,$$
$$F_5=(D_3+D_4 \psi-2 \alpha D_3),$$
$$F_6=(D_4+D_3 \psi-2 \alpha D_4),$$
$$F_7=(D_4\psi R^2 \psi_1-D_3 R^2 \psi_2),$$
$$F_8=(D_3\psi R^2\psi_1-D_4 R^2\psi_2)),$$
$$F_9=(D_3-D_4 \psi),~F_10=(D_4-D_3 \psi).$$

We, thus, have provided three different class of solutions for $0 \leq \alpha <1$ (Case I), $\alpha =1$ (Case II) and $ \alpha > 1$ (Case III). We note that all the solutions are regular. One, however, needs to examine the physical viability of the solutions which can be analyzed by utilizing the junction conditions and systematically fixing the values of the model parameters. An interesting feature of the class of solutions is that they provide a mechanism to examine the impact of anisotropy on the physical properties of a relativistic star simply by using the parameter $\alpha$ as an `anisotropic switch'.

\section{Boundary conditions}\label{sec3}
The spacetime metric (\ref{1}) must be matched to the Schwarzschild exterior metric
\begin{eqnarray}\label{EMetric}
    ds^{2}&=&\left(1-\frac{2M}{r}\right)dt^{2}-\left(1-\frac{2M}{r}\right)^{-1}dr^{2}\nonumber\\
		&&-r^{2}\left(d\theta^{2}+\sin^{2}\theta d\phi^{2} \right),
\end{eqnarray}
at the boundary of the star $r=b$ where $M=m(b)$ is the total mass of the star. This implies 
\begin{equation}
e^{2\nu(r=b)}= \left(1-\frac{2 M}{b}\right),\label{bc1}
\end{equation}
\begin{equation}
Z(r=b)= \left(1-\frac{2 M}{b}\right).\label{bc2}
\end{equation}
The radius $b$ is defined as the surface where the radial pressure drops to zero ($p_r(r=b)=0$). The above boundary conditions determine the constants of the solutions which are obtained for three different cases below:

\subsection{Case I:} 
The matching conditions yield
\begin{eqnarray}
1-\frac{2 M}{b}&=&A^2\left[\left(D_1 - D_2\xi_b\right)\cos\xi_b\right.\nonumber\\
&&\left.+ \left(D_2 + D_1\xi_b\right)\sin\xi_b\right]^2,\label{bc5}\\
\frac{1}{1+\frac{b^2}{R^2}}&=&A^2 \left[(D_1 - D_2 \xi_b) \cos\xi_b\right.\nonumber\\
&&\left. + (D_2 + D_1 \xi_b)\sin \xi_b \right]^2,\label{bc6}
\end{eqnarray}
where $$\xi_b=\sqrt{(1-\alpha)\left(1+\frac{b^2}{R^2}\right)}$$ is the value of $\xi$ at $r=b$.\\
On imposing the condition $p_r(r=b)=0$, we get
\begin{eqnarray}
(D_1+D_2 \xi_b-2 \alpha D_1) \cos \xi_b\nonumber\\
+ (D_2-D_1 \xi_b-2 \alpha D_2)\sin \xi_b=0, \label{bc7}
\end{eqnarray}
which can be written as
\begin{equation}
\frac{D_2}{D_1}=\frac{(2\alpha-1)+\xi_b Tan \xi_b}{(1-2\alpha)Tan \xi_b+\xi_b}. \label{bc8}
\end{equation}

Combining Eqs.~(\ref{bc6}) and Eq.~(\ref{bc8}), we obtain 
\begin{eqnarray}
D_1&=&\frac{[\xi_b \cos \xi_b+(1-2\alpha)\sin \xi_b]}{2 A \xi_b^2 \sqrt{1-\alpha}},\label{bc10}\\
D_2&=&\frac{\xi_b \sin \xi_b-(1-2\alpha)\cos \xi_b]}{2 A \xi_b^2 \sqrt{1-\alpha}}.\label{bc11}
\end{eqnarray}

\subsection{Case II:}
The matching conditions yield
\begin{equation}
1-\frac{2 M}{b}=A^2 \left[a_1+b_1(1+\frac{b^2}{R^2})^{3/2}\right]^2, \label{bc5e}
\end{equation}
\begin{equation}
\frac{1}{1+\frac{b^2}{R^2}}=A^2 \left[a_1+b_1(1+\frac{b^2}{R^2})^{3/2}\right]^2.\label{bc6e}
\end{equation}

On imposing the condition $p_r(r=b)=0$, we get
\begin{equation}
\left[a_1+b_1(1+\frac{b^2}{R^2})^{3/2}\right]=6b_1\sqrt{1+\frac{b^2}{R^2}}. \label{bc7e}
\end{equation}
Combining Eq.~(\ref{bc6e}) and (\ref{bc7e}), we obtain
\begin{eqnarray}
b_1&=&\frac{1}{6A(1+\frac{b^2}{R^2})},\label{bc8e}\\
a_1&=&\frac{5-\frac{b^2}{R^2}}{6A \sqrt{(1+\frac{b^2}{R^2})}}.\label{bc9e}
\end{eqnarray}

\subsection{Case III:}
The matching conditions yield
\begin{eqnarray}
1-\frac{2 M}{b}&=&A^2 \left[\left(D_4 - D_3\psi_b\right)\sinh\psi_b\right.\nonumber\\
&&\left.+\left(D_3 - D_4\psi_b\right)\cosh\psi_b\right]^2,\label{bc5g}\\
\frac{1}{1+\frac{b^2}{R^2}}&=&A^2 \left[ (D_4 - D_3
\psi_b) \sinh\psi_b\right.\nonumber\\
 &&\left.+ (D_3 - D_4\psi_b)\cosh\psi_b\right]^2.\label{bc6g}
\end{eqnarray}
On imposing the condition $p_r(r=b)=0$, we get
\begin{eqnarray}
(D_3+D_4 \psi_b-2 \alpha D_3) \cos \psi_b\nonumber\\
+ (D_4-D_3 \psi_b-2 \alpha D_4)\sin \psi_b=0, \label{bc7g}
\end{eqnarray}
where $$\psi_b=\sqrt{(\alpha-1)\left(1+\frac{b^2}{R^2}\right)},$$ is the value of $\psi$ at $r=b$.
Eq.~(\ref{bc7g}) can be written as
\begin{equation}
\frac{D_4}{D_3}=\frac{(2\alpha-1)\cosh \psi_b-\psi_b \sinh \psi_b}{\psi_b \cosh \psi_b+(1-2\alpha)\sinh \psi_b}, \label{bc8g}
\end{equation}
which together with Eq.~(\ref{bc6g}) determine the constants as
\begin{equation}
D_3=\frac{\psi_b \cosh \psi_b+(1-2\alpha)\sinh \psi_b}{2 A \psi_b^2 \sqrt{\alpha-1}},\label{bc9g}
\end{equation}
\begin{equation}
D_4=\frac{(2\alpha-1) \cosh \psi_b-\psi_b \sinh\psi_b}{2 A \psi_b^2 \sqrt{\alpha-1}}.\label{bc10g}
\end{equation}

It is important to note that, in all the three cases, the parameter $A$ can be absorbed by redefining the constants. 

\section{Physical analysis}\label{sec4}
Following \cite{Delgaty98}, we demand that a physically viable stellar model should satisfy the following conditions throughout the stellar interior: \\
\noindent (i) Density, radial pressure and transverse pressure should be positive and finite throughout the star i.e., $ \rho \geq 0$, ~~~ $p_r \geq 0 $, ~~~ $p_t \geq 0 $; \\
\noindent (ii)  Fulfillment of the strong energy condition within the star i.e., $ \rho -  p_r- 2 p_t   \geq 0 $;\\
\noindent (iii) The energy density should be  a decreasing function of $r$ i.e., $ \frac{d\rho}{dr} < 0 $;\\
\noindent (iv) The gradient of the pressure must be negative inside the stellar configuration i.e., $\frac{dp_r}{dr} < 0 $, ~~~ $\frac{dp_t}{dr} < 0 $;\\
\noindent (v) The speed of sound must be smaller than the speed of light in the stellar interior i.e., $ 0 \leq \frac{dp_r}{d\rho} \leq 1 $, $ 0 \leq \frac{dp_t}{d\rho} \leq 1 $;\\
\noindent (vi) The adiabatic index  
\begin{equation}
\Gamma=\frac{\rho+p}{p}\frac{dp}{d\rho},\nonumber
\end{equation} 
should be greater than $4/3$ for stability of the configuration (\cite{hm}). 

These requirements can be tested for all the three class of solutions provided in this paper. However, since the solution within $0 \leq \alpha < 1$ (Case I) has an isotropic counterpart it provides a simple tool to analyze the impact of anisotropy onto the system. According, we examine the physical viability of this particular solution in this work.  

\begin{figure}
\includegraphics[width=0.45\textwidth]{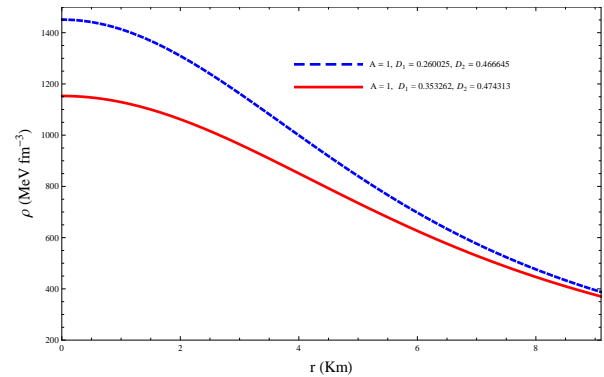}
\caption{Variation of density.}
\label{fig1}
\end{figure}

\begin{figure}
\includegraphics[width=0.45\textwidth]{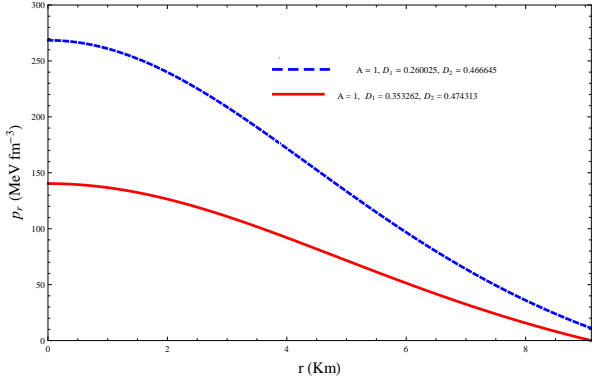}
\caption{Variation of radial pressure.}
\label{fig2}
\end{figure}

\begin{figure}
\includegraphics[width=0.45\textwidth]{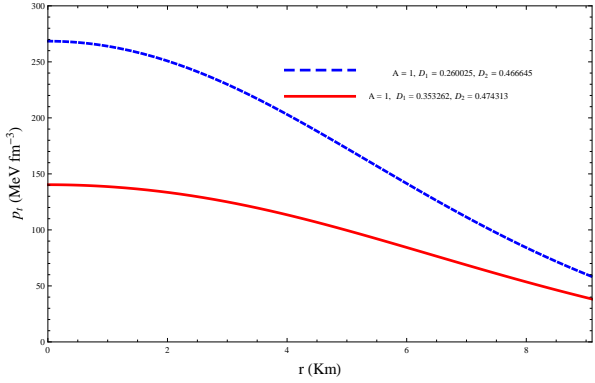}
\caption{Variation of transverse pressure.}
\label{fig3}
\end{figure}

\begin{figure}
\includegraphics[width=0.45\textwidth]{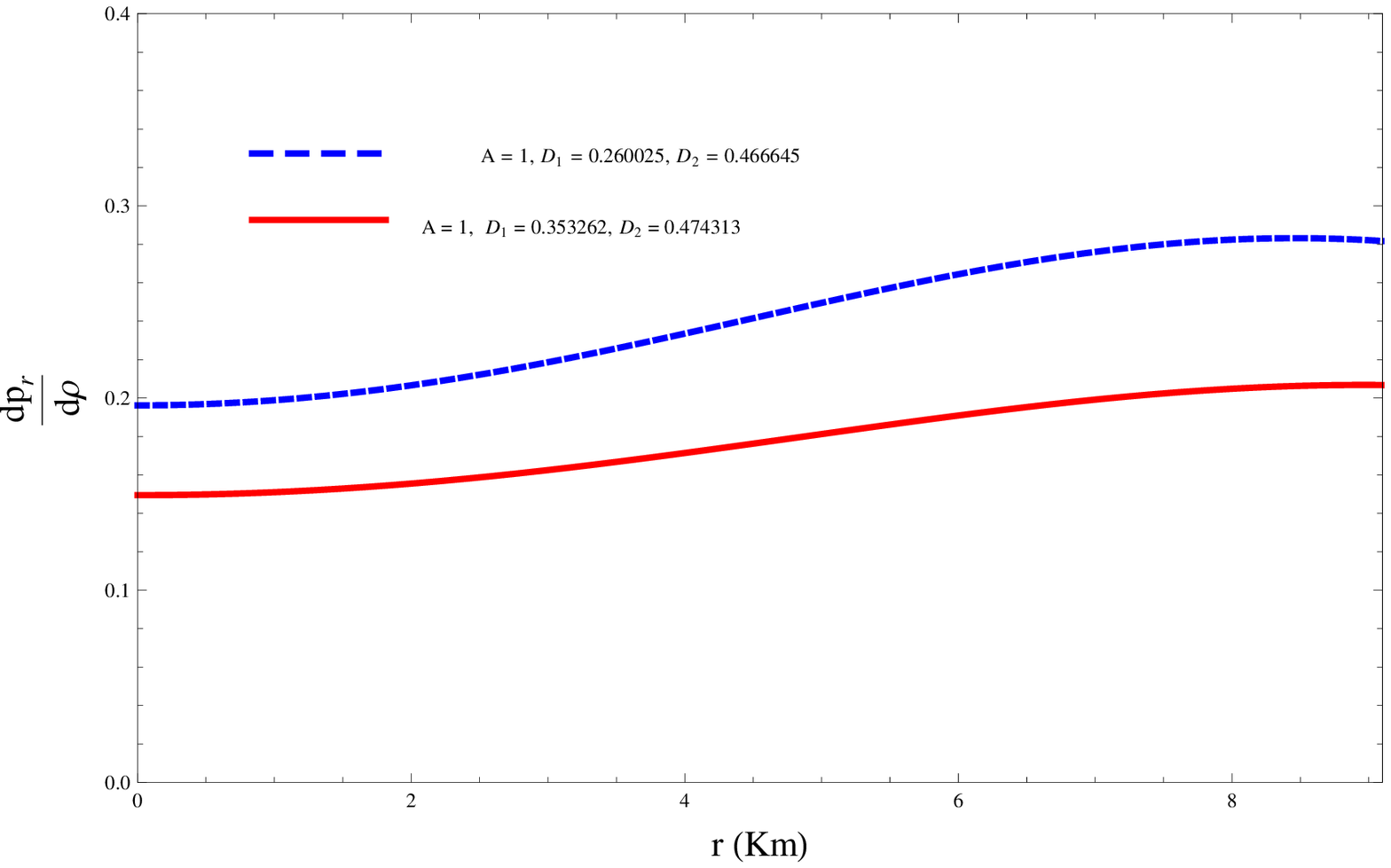}
\caption{Variation of $dp_{r}/d \rho$ with radial parameter. }
\label{fig4}
\end{figure}

\begin{figure}
\includegraphics[width=0.45\textwidth]{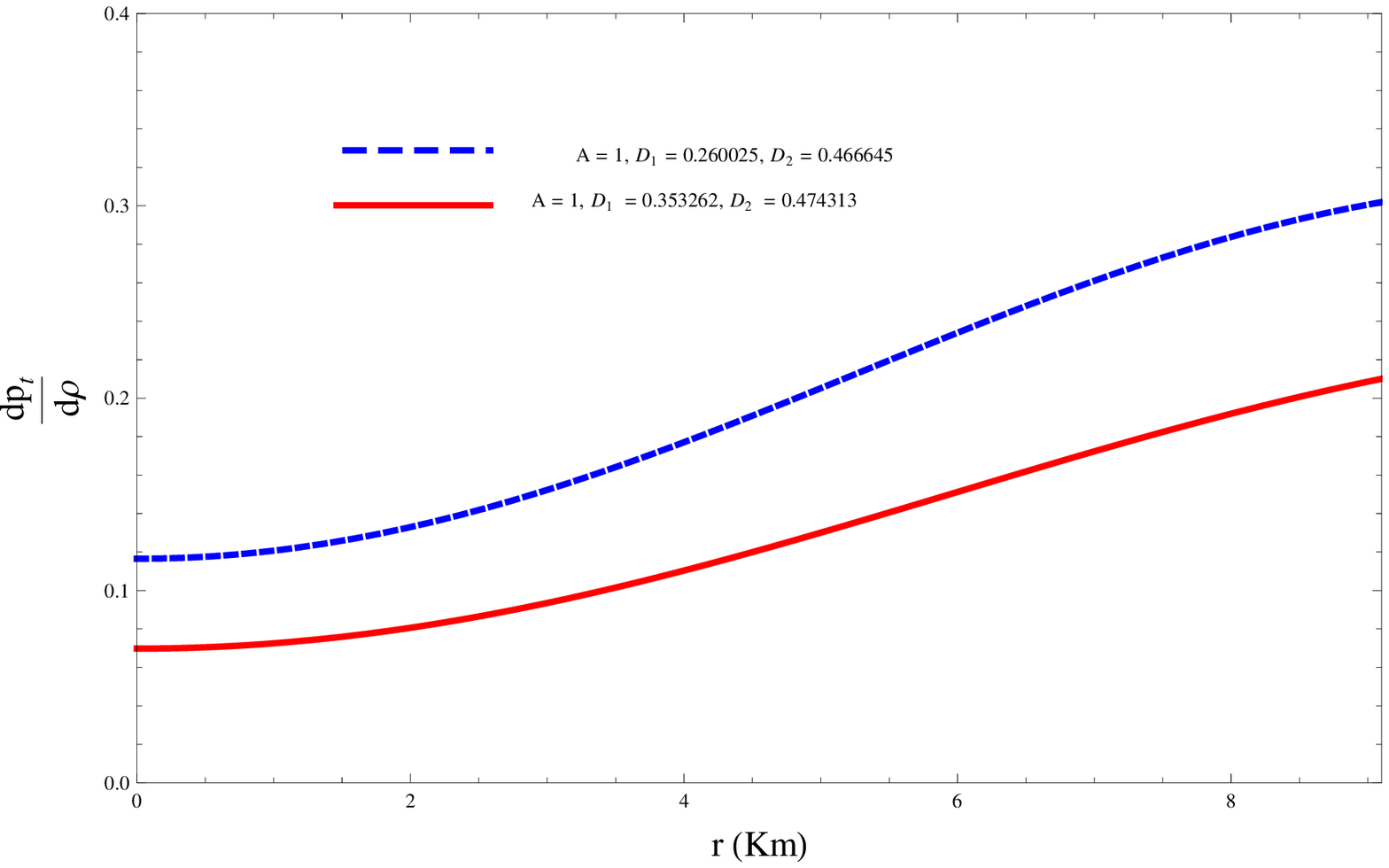}
\caption{Variation of $dp_{t}/d \rho$ with radial parameter.}
\label{fig5}
\end{figure}

\begin{figure}
\includegraphics[width=0.45\textwidth]{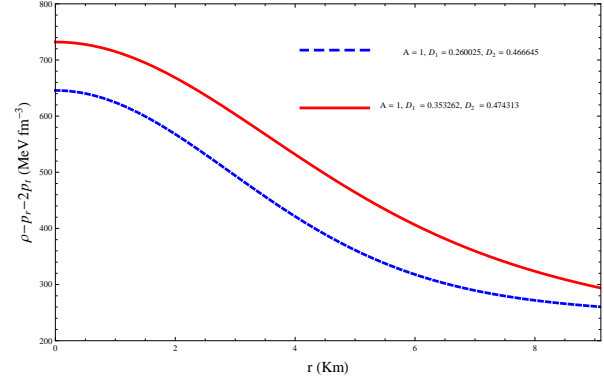}
\caption{Variation of $\rho -p_r -2p_t$ with radial parameter.}
\label{fig6}
\end{figure}

\begin{figure}
\includegraphics[width=0.45\textwidth]{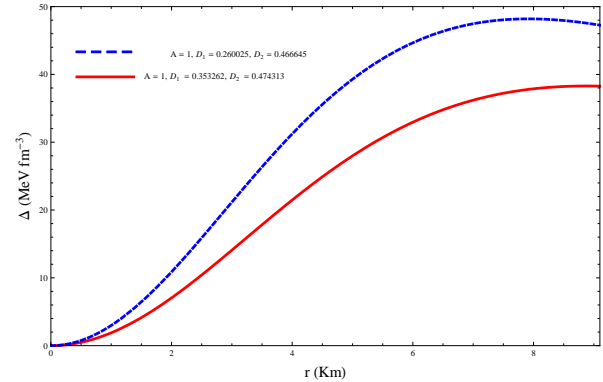}
\caption{Variation of anisotropy.}
\label{fig7}
\end{figure}

\begin{figure}
\includegraphics[width=0.45\textwidth]{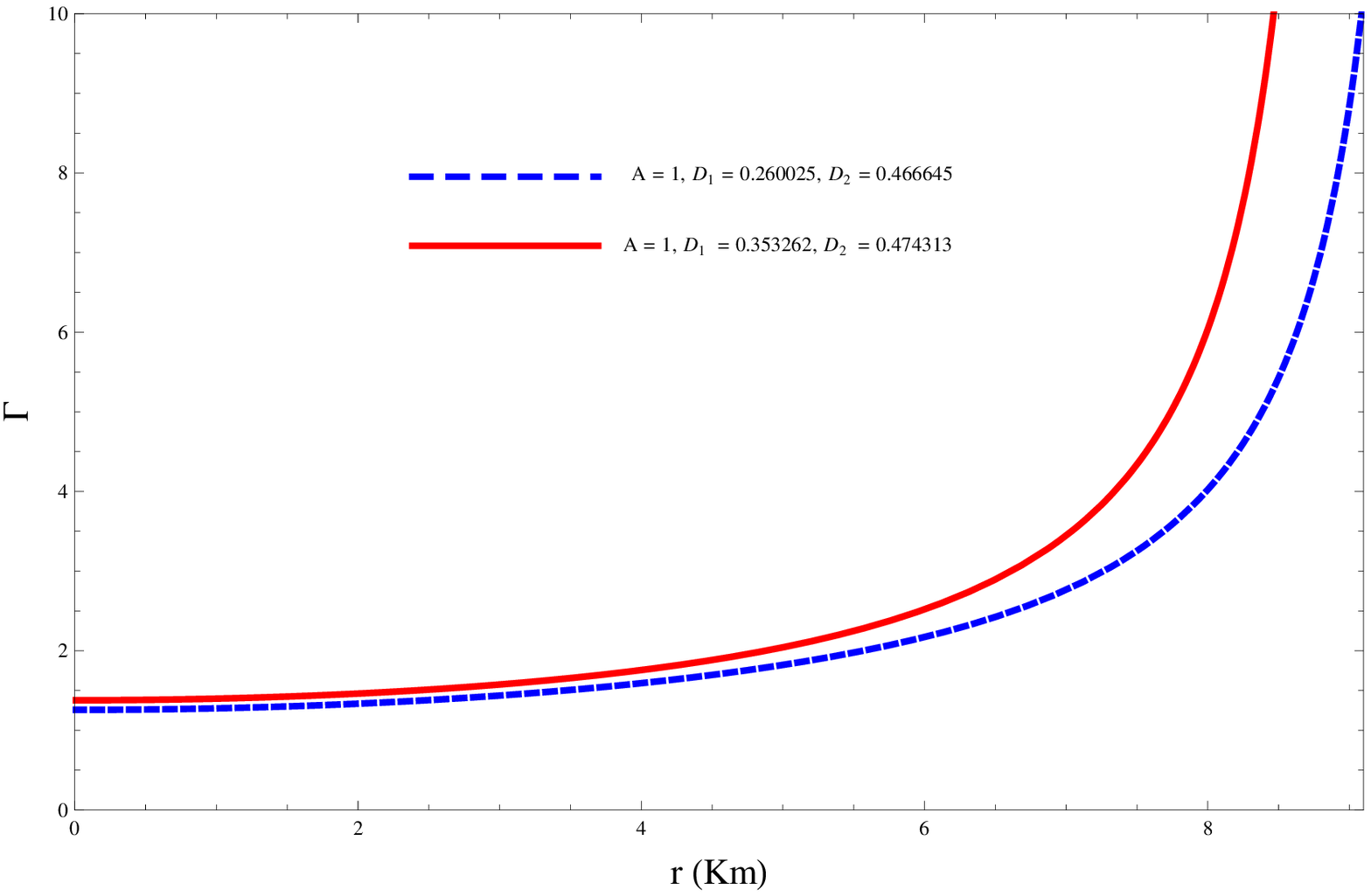}
\caption{Variation adiabatic index.}
\label{fig8}
\end{figure}

To examine whether our class of solutions satisfy all the requirements of a relistic star, we consider the data available from the pulsars $4U1820-30$ and $PSR J1614-2230$ whose estimated masses and radii are $M = 1.58 ~M_{\odot}$, $b = 9.1~km$ (\cite{Guver10}) and  $M = 1.97~ M_{\odot}$, $b = 9.69~ km$ (\cite{Demorest10}), respectively. Using these values as inputs for a given anisotropic configuration (we have assumed $\alpha=0.4$), the boundary conditions have been utilized to determine the constants which have been compiled in Table~(\ref{tab1}). Making use of these values and also plugging in $8\pi G$ and $c$ at appropriate places, all the relevant physically meaningful quantities have been plotted in Fig.~(\ref{fig1}) - (\ref{fig8}). In the plots, the red and dashed blue colours have been used to distinguish the pulsar $4U1820 - 30$ and $PSR J1614 - 2230$, respectively. Fig.~(\ref{fig1}) shows that for both the pulsars the density decreases from its maximum value at the centre towards the boundary. In Fig.~(\ref{fig2})-(\ref{fig3}), radial variation of two pressures has been plotted. In Fig.~(\ref{fig4})-(\ref{fig5}), radial variation of sound speed in the radial and transverse directions have been shown which confirms that the causality condition is not violated throughout the interior of the star. Fig.~(\ref{fig6}) shows that the strong energy condition is also not violated throughout the stellar interior. Radial variation of the anisotropic parameter has been shown in Fig.~(\ref{fig7}). Variation of the adiabatic index has been plotted in Fig.~(\ref{fig8}) which clearly indicates stability of the configurations. 
\begin{center}
\begin{table*}
\small
\caption{\label{tab1} Model parameters.}
\begin{tabular}{|c|c|c|c|c|c|c|c|    c|c|}\hline
$ Pulsar $ & $Mass (M_{\odot})$ & $ Radius~$(Km)  & $ \alpha $ & $ A $ & $D_1$ & $D_2$ & $R~$(km) \\ \hline
$4U1820-30$ & $1.58$  & $9.1$ &  $0.4$ & $1$ & $0.353262$ & $0.474313$ & $8.88064$ \\ \hline
                                                        
$PSR J1614-2230$ & $1.97$ & $9.69$ &  $0.4$ & $1$ & $0.260025$ & $0.466645$ & $7.91611$ \\ \hline
                                                        
$Cen X-3$   & $1.49$ & $9.178$  & $0$ & $1$ & $0.322637$ & $0.306689$ & $9.57351$ \\
                                                          
              &        &          &    $0.5$ & $1$ & $0.402356$ & $0.599327$ & $9.57351$  \\ \hline

\end{tabular}
\end{table*}
\end{center}

\begin{figure}
\includegraphics[width=0.45\textwidth]{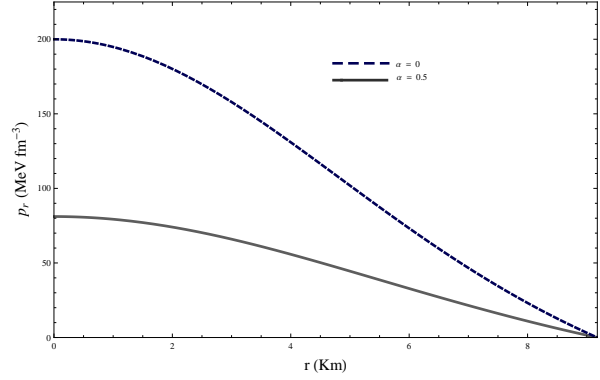}
\caption{Comparison of radial pressure for different $\alpha$.}
\label{fig9}
\end{figure}

\begin{figure}
\includegraphics[width=0.45\textwidth]{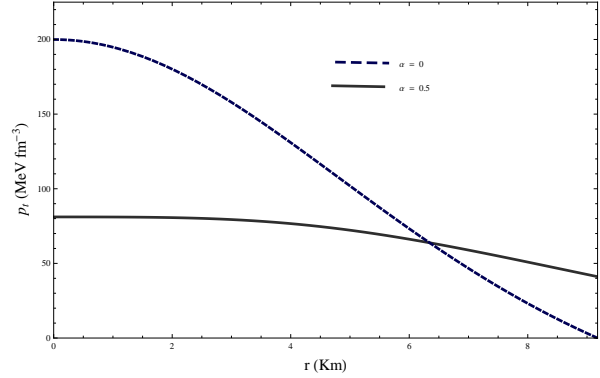}
\caption{Comparison of tangential pressure for different $\alpha$.}
\label{fig10}
\end{figure}

\begin{figure}
\includegraphics[width=0.45\textwidth]{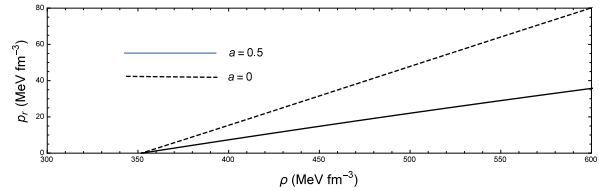}
\caption{Comparison of EOS for different $\alpha$.}
\label{fig12}
\end{figure}

To examine the impact of anisotropy explicitly, let us now consider two hypothetical situations. Assuming that the pulsar $Cen X-3$ has either an isotropic ($\alpha=0$) or an anisotropic matter distribution (for which we choose $\alpha=0.5$), we have analyzed the behaviour of the physical quantities for the two possible scenarios. By fixing the model parameters for an estimated mass $M = 1.49 ~M_{\odot}$ and radius $b = 9.178~km$ of the pulsar (\cite{Rawls11}), the variation of different physical quantities has been analyzed for the two cases. We note that, even though for a given mass and radius the density profile remains the same, we observe that to accommodate anisotropy the pressure decreases near the centre as shown in Fig.~(\ref{fig9})-(\ref{fig10}). Obviously, the lowering of pressure for a given density profile points towards a change in EOS when anisotropy develops in the matter composition.  This has been demonstrated in Fig.~(\ref{fig12}) where it has been shown that anisotropy might provide a comparatively softer EOS. However, since the source of anisotropy is not known in our construction, it is difficult to draw any conclusive physical inference based on our observation.

\begin{figure}
\includegraphics[width=0.45\textwidth]{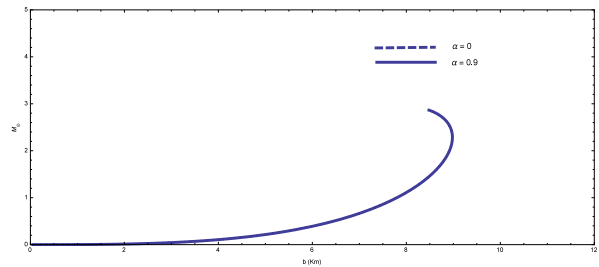}
\caption{Impact of anisotropy on the mass-radius relationship. For a given surface density ($\rho_s = 7.5\times 10^{14}~$gm~cm$^{-3}$), the dotted curve corresponds to $\alpha=0$ case and the solid curve corresponds to $\alpha=0$ case. The overlapped nature of the plots show that the mass-radius relationship for a given density remains unaffected by the presence/absence of anisotropy.}
\label{fig11}
\end{figure}

We have also analyzed the mass-radius ($M-b$) relationship of the model for different values of the anisotropic parameter (e.g., $\alpha = 0,~0.9$). It is noteworthy that for a given surface density (for numerical calculation, we have assumed surface density $\rho_s = 7.5\times 10^{14}~$gm~cm$^{-3}$), pressure anisotropy seems to have no impact on the overall mass-radius relationship of the star as can be seen in Fig.~(\ref{fig11}). Similar results were obtained by \cite{Sunzu14} for some  specific set of model parameters in their paper. It is noteworthy that in our model density remains unaffected even when anisotropy is induced into the system. Consequently, for a given central and/or surface density, an external observer sees no change in its mass-radius relationship. In a separate study (\cite{Ratanpal17}) it has been observed that in the presence of an electric field  a stellar configuration tends to accommodate more mass. However, it should be stressed that incorporation of an electromagnetic field changes the exterior spacetime from Schwarzschild to Reissner-Nordstr\"om whereas in our case the exterior spacetime remains unaltered. Nevertheless, whether exhibition of such distinct features is generic in nature can be understood by initiating further studies in this direction for a wider class of anisotropic stellar solutions.

\section{Concluding remarks}\label{sec5}

The \cite{Finch89} model has been extensively used to study different types of matter distributions and consequently, a plethora of solutions have been obtained by many investigators. What is interesting about the different family of solutions is that it is possible to obtain a similar class of solutions with different physical motivations vis-a-vis matter compositions. In this paper, we have generated new class of solutions for a wide range of anisotropic parameter. In the context of our developed model, we would like to point out the following:
\begin{itemize}

\item The Finch-Skea solution has been generalized by many to study charged fluid distributions. While a charge-fluid distribution has an anisotropic interpretation, an anisotropic matter distribution is not necessarily charged. Anisotropy, naturally, demands a separate analysis.

\item In some papers, solutions have been obtained for matter distributions which are charged as well as anisotropic in nature. While the physical motivation for such propositions cannot be ignored, it should be noted that from a mathematical point of view, an electromagnetic field can always be absorbed in a more general anisotropic term, particularly if the electric field intensity $E^2$ follows the same radial fall-off behaviour as that of the anisotropic parameter $\Delta$.

\item  It is worthwhile to note that our class of solutions is similar to the family of solutions obtained by \cite{Maharaj16} even though the methods adopted are different. \cite{Maharaj16} have generated solutions to the master differential equation in terms of the Bessel functions which have later been expressed in closed forms for some specific range of model parameters defining the electromagnetic field and anisotropy. In our work, we have utilized the Frobenius method to solve the master differential equation and showed that similar class of solutions could be obtained by terminating the series solution for a specific range of anisotropy.

\item It is noteworthy that our class of solutions do not contain the case for which $p_r > p_t$.

\item As far as the impact of anisotropy on the stellar behaviour is concerned, we note that for a given mass and radius, the inclusion of anisotropy decreases the pressure near the central region of the star. However, for a given surface density, pressure anisotropy appears to have no role on the mass-radius relationship of the star.

\item Finally, the constant $A$ appearing in the \cite{Durga} transformation equations seems to have no role in the resultant configuration. In fact, in this construction, the parameter $A$ can be absorbed by re-defining the other constants appearing in the metric potentials. While in the \cite{Durga} paper, $A$ can be fixed from the boundary conditions, when additional degrees of freedom is introduced in the form of anisotropy and/or charge, the nature of the solutions are such that $A$ remains as a free parameter in this formulation.
\end{itemize}

\section{Acknowledgements}
\noindent The authors would like to thank the anonymous referees for his very useful comments and suggestions. The work of RS is supported by the MRP research grant no. F.PSW-195/15-16 (ERO) of the UGC, Govt. of India. RS also gratefully acknowledges support from the Inter-University Centre for Astronomy and Astrophysics (IUCAA), Pune, India, under its Visiting Research Associateship Programme. SD is thankful to IUCAA where a part of this work was carried out.


\begin{thebibliography}{99}

\bibitem[\protect \citeauthoryear{Finch and Skea}{1989}]{Finch89} Finch M. R. and Skea J. E. F., {\it Class. Quantum Grav.} {\bf6} (1989) 467.
\bibitem[\protect \citeauthoryear{Duorah and Ray}{1987}]{Duorah87} Duorah H. L. and Ray R., {\it Class. Quantum Gravity} {\bf4} (1987) 1691.
\bibitem[\protect \citeauthoryear{Delgaty and Lake}{1998}]{Delgaty98} Delgaty M. S. R. and Lake K., {\it Comput. Phys. Commun.} {\bf115} (1998) 395; \\
doi: {\url{http://dx.doi.org/10.1016/s0010-4655(98)00130-1}}.
\bibitem[\protect \citeauthoryear{Finch and Skea}{1998}]{Finch98} Finch M. R. and Skea J. E. F.,\\ {\em http://edradour.symbcomp.uerj.br/pubs.html} (1998).
\bibitem[\protect \citeauthoryear{Maharaj {\em et al}}{2016}]{Maharaj16} Maharaj S. D., Matondo D. K. and  Takisa  P. M., {\it Int. J. Mod. Phys. D} {\bf26} (2016)  1750014.
\bibitem[\protect \citeauthoryear{Kileba {\em et al}}{2017}]{Kileba17} Matondo D. K., Takisa P. M., Maharaj S. D. and  Ray S., {\it Astrophys. Space Sci.} {\bf362} (2017) 186.
\bibitem[\protect \citeauthoryear{Sharma and Das}{2013}]{Sharma2013} Sharma R. and Das S., {\it J. Gravit.} {\bf 2013} (2013) 659605.
\bibitem[\protect \citeauthoryear{Ratanpal {\em et al}}{2017}]{Ratanpal17} Ratanpal B. S., Pandya D. M., Sharma R. and Das S., {\it Astrophys. Space Sci.} {\bf 362} (2017) 82.  
\bibitem[\protect \citeauthoryear{Hansraj and Maharaj}{2006}]{Hansraj06} Hansraj S. and Maharaj S. D., {\it Int. J. Mod. Phys. D.} {\bf8} (2006) 1311.
\bibitem[\protect \citeauthoryear{Pandya {\em et al}}{2015}]{Pandya15} Pandya D. M., Thomas V. O. and Sharma R., {\it Astrophys. Space Sci.} {\bf 356} (2015) 285.
\bibitem[\protect \citeauthoryear{Sharma and Ratanpal}{2013}]{Sharma13} Sharma R. and Ratanpal B. S., {\it Int. J. Mod. Phys. D} {\bf13} (2013) 1350074.
\bibitem[\protect \citeauthoryear{Bhar {\em et al}}{2014}]{Bhar14} Bhar P., Rahaman. F., Biswas. R. and Fatima H. I., {\it Commun. Theor. Phys.} {\bf 62} (2014) 221.
\bibitem[\protect \citeauthoryear{Ba\~{n}ados {\it et al}}{1992}]{BTZ} Ba$\tilde{n}$ados M., Teitelboim C. and Zanelli J., {\it Phys. Rev. Lett.} {\bf69} (1992) 1849.
\bibitem[\protect \citeauthoryear{Chodos {\em et al}}{1974}]{MIT} Chodos A., Jaffe R. L., Johnson K., Thorn C. B. and Weisskopf V. F., {\it Phys. Rev. D}{\bf9} (1974) 3471.
\bibitem[\protect \citeauthoryear{Banerjee {\em et al}}{2013}]{Ayan13} Banerjee A., Rahaman F., Jotania K., Sharma R. and Karar I., {\it Gen. Revativ. Grav.} {\bf45} (2013) 717.
\bibitem[\protect \citeauthoryear{Hansraj {\em et al}}{2017}]{Hansraj17} Hansraj S., Goswami R., Ellis G. and Mkhize N., {\it Phys. Rev. D}{\bf 96} (2017) 044016.
\bibitem[\protect \citeauthoryear{Hansraj }{2016}]{Hansraj16} Hansraj S., {\it Eur. Phys. J. C.} {\bf 77} (2017) 557.
\bibitem[\protect \citeauthoryear{Dadhich {\em et al}}{2016}]{Dadhich16} Dadich N., Hansraj S. and Chilambwe B., {\it Int. J. Mod. Phys. D} {\bf26} (2017) 1750056.
\bibitem[\protect \citeauthoryear{Molina {\em et al}}{2017}]{Molina17}  Molina A., Dadich N. and Khugaev A., {\it Gen. Rel. Gravit} {\bf49} (2017) 96. 
\bibitem[\protect \citeauthoryear{Ruderman}{1972}]{Ruderman72} Ruderman R., {\it Astro. Astrophys.} {\bf10} (1972) 427.
\bibitem[\protect \citeauthoryear{Canuto}{1974}]{Canuto74} Canuto V., {\it Annu. Rev. Astron. Astrophys.} {\bf12} (1974) 167.
\bibitem[\protect \citeauthoryear{Bowers and Liang}{1974}]{Bowers74} Bowers R. and Liang E., {\it Astrophys. J.} {\bf188} (1974) 657.
\bibitem[\protect \citeauthoryear{Kippenhahn and Weigert}{1990}]{Kippen} Kippenhahn R. and  Weigert A., Stellar Structure and Evolution. {\it Springer-Verlag, Berlin} (1990).
\bibitem[\protect \citeauthoryear{Weber}{1999}]{Weber} Weber F., Pulsars as Astrophysical Observatories for Nuclear and Particle Physics. {\it IOP Publishing, Bristol} (1999).
\bibitem[\protect \citeauthoryear{Usov}{2004}]{Usov} Usov V. V., {\it Phys. Rev. D} {\bf70} (2004) 067301. 
\bibitem[\protect \citeauthoryear{Sokolov}{1980}]{Sokolov} Sokolov A. I., {\it JETP} {\bf79} (1980) 1137.
\bibitem[\protect \citeauthoryear{Sawyer}{1972}]{Sawyer} Sawyer R. F., {\it Phys. Rev. Lett.} {\bf 29} (1972) 382.
\bibitem[\protect \citeauthoryear{Schunch}{2003}]{Schunck} Schunck F. E. and Mielke E. W., {\it Class. Quantum Grav.} {\bf 20} (2003) R301.
\bibitem[\protect \citeauthoryear{Morris and Thorne}{1988}]{Morris} Morris M. S. and Thorne K. S., {\it Am. J. Phys.} {\bf 56} (1988) 395.      
\bibitem[\protect \citeauthoryear{Cattoen {\em et al}}{2005}]{Cattoen} Cattoen C., Faber T. and Visser M., {\it Class. Quantum Grav.} {\bf 22} (2005) 4189.
\bibitem[\protect \citeauthoryear{DeBenedictis {\em et al}}{2006}]{DeBend} DeBenedictis A., Horvat D., Ilijic S., Kloster S. and  Viswanathan K., {\it Class. Quantum Grav.} {\bf 23} (2006) 2303.
\bibitem[\protect \citeauthoryear{Ivanov}{2010}]{Ivanov10} Ivanov B. V., {\it Int. J. Theor. Phys.} {\bf 49} (2010) 1236.
\bibitem[\protect \citeauthoryear{Ivanov}{2017}]{Ivanov17} Ivanov B. V., eprint arXiv: 1708.07971 [gr-qc].
\bibitem[\protect \citeauthoryear{Tikekar and Jotania}{2007}]{Tikekar07} Tikekar R. and Jotania K., {\it Pramana-j. of physics} {\bf68} (2007) 397.
\bibitem[\protect \citeauthoryear{Durgapal and Bannerji}{1983}]{Durga} Durgapal M. C. and Bannerji R., {\it Phys. Rev.} {\bf27} (1983) 328.
\bibitem[\protect \citeauthoryear{Heintzmann and Hilebrandt }{2010}]{hm} Heintzmann H. and  Hillebrandt W., {\it Astron. Astrophys.} {\bf38} (1975) 51.
\bibitem[\protect \citeauthoryear{G\"{u}ver {\em et al} }{2010}]{Guver10} G\"{u}ver T., \"{O}zel F., Cabrera-Lavers A. and Wroblewski P., {\it ApJ.} {\bf712} (2010) 964. 
\bibitem[\protect \citeauthoryear{Demorest {\em et al}}{2010}]{Demorest10} Demorest P. B., Pennucci T., Ranson S. M., Rpberts M. S. E. and Hessels J. W. T., {\it Nat.} {\bf467} (2010) 1081.
\bibitem[\protect \citeauthoryear{Rawls {\em et al}}{2011}]{Rawls11} Rawls M. L., Orosz J. A., McClintock J. E., Torres M. A. P., Baliyn C. D. and Buxton M. M., {\it ApJ.} {\bf730} (2011) 25.
\bibitem[\protect \citeauthoryear{Sunzu {\em et al}}{2014}]{Sunzu14} Sunzu J. M., Maharaj S. D. and Ray S., {\it Astrophys. Space Sci.} {\bf352} (2014)  719.



\end{thebibliography}
\end{document}